\documentclass[finallayout,an,fleqn]{w-art}
\usepackage{times}
\usepackage{w-thm}
\theoremstyle{plain}

\theoremstyle{definition}

\usepackage[]{graphicx}
\chardef\bslash=`\\ 

\begin{document}

\DOIsuffix{theDOIsuffix}
\Volume{324}
\Issue{S1}
\Copyrightissue{S1}
\Month{01}
\Year{2003}
\pagespan{3}{}
\Receiveddate{7 November 2002}
\Reviseddate{30 November 2002}
\Accepteddate{2 December 2002}
\Dateposted{3 December 2002}



\title[Photoionization and photodissociation in Sgr B2]
{Extended photoionization and photodissociation in Sgr B2}


\author[Goicoechea]
{J.R. Goicoechea\footnote{Corresponding
     author: e-mail: {\sf javier@isis.iem.csic.es}, Phone: +0034\,91 561 68 00 ,
     Fax: +0034\,91 564 55 57}\inst{1}} \address[\inst{1}]
     {Departamento de Astrof\'{\i}sica Molecular e Infrarroja,
     IEM/CSIC, Serrano 121, E--28006 Madrid, Spain}
\author[Rodr\'{\i}guez-Fern\'andez]
{N.J. Rodr\'{\i}guez-Fern\'andez
\footnote{e-mail: {\sf Nemesio.Rodriguez-Fernandez@obspm.fr}}\inst{2}}\address[\inst{2}]
{Observatoire de Paris - LERMA. 61, Av. de l'Observatoire, 75014 Paris, France}
\author[Cernicharo]
{J. Cernicharo\footnote{e-mail: {\sf cerni@astro.iem.csic.es}}\inst{1}}

\begin{abstract}

We present large scale $9'\times27'$ (25 pc $\times$ 70 pc) 
far--IR  observations of the Sgr B2 complex using the spectrometers
on board the \textit{Infrared Space Observatory}$^1$ (ISO).
The far--IR spectra are dominated by the strong continuum emission
of dust and by  the fine structure lines of high excitation potential 
ions (NII, NIII and OIII) and those of neutral or weakly ionized atoms
(OI and CII). The line emission has revealed a very extended component
of ionized gas.
The study of the NIII 57 $\mu$m$/$NII 122 $\mu$m and
OIII 52$\mu$m $/$88$\mu$m line intensity ratios show that
the ionized gas has a density of n$_e$$\simeq$10$^{2-3}$ cm$^{-3}$ while 
the ionizing radiation can be characterized by a diluted but hard
continuum, with effective temperatures of $\sim$35000 K.

Photoionization models show that the total number of Lyman
photons needed to explain such an extended component is approximately
equal to that of the HII regions in Sgr B2(N) and (M) condensations.
We propose that the  inhomogeneous and clumpy structure of the cloud 
allows the radiation to reach large distances through the envelope. 
Therefore, photodissociation regions (PDRs) can be numerous at the 
interface of the ionized and the neutral gas.
The analysis of the OI (63 and 145 $\mu$m) and CII (158 $\mu$m) lines 
indicates an incident far--UV field (G$_0$, in units of the local interstellar
radiation field) of 10$^{3-4}$ and a H density of  10$^{3-4}$ cm$^{-3}$ in such PDRs. 
We conclude that extended photoionization 
and photodissociation are also taking place in Sgr B2 in addition 
to more established phenomena such as widespread low--velocity shocks.

\end{abstract}

\maketitle                   





\section{Introduction}
  
\begin{figure}[t]
\centerline{\includegraphics[width=12cm]{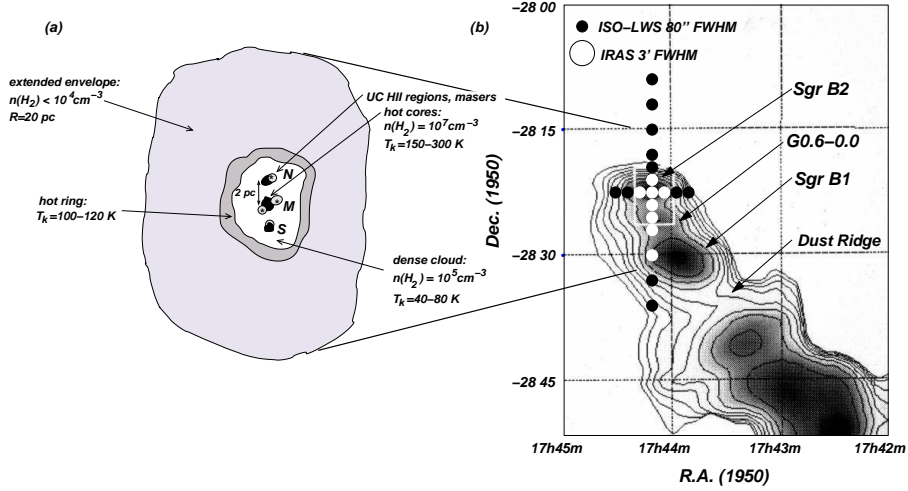}}
\caption{$Right:$ Large scale IRAS image at 60 $\mu$m (Gordon et al. 1993) and
ISO target positions across Sgr B2 region.
$Left:$ Sketch showing the different structures and components in the
Sgr B2 complex. Hot cores are shown black shaded and HII regions
are the structures enclosing the stars. 
(Adapted from H\"{u}ttemeister et al. 1995)}
\label{fig:1}
\end{figure}

The Sgr B2 complex represents an interesting burst of massive
star formation in the inner 400 pc of the Galaxy (that we refer
as the Galactic Center region [GC])
and may be representative of other active nuclei. 
Large scale continuum emission studies show that Sgr B2 is the brightest 
emission and the most massive cloud of 
the region ($\simeq$10$^7$ M$_{\odot}$; Lis \& Goldsmith 1989). The main
signposts of star activity are located within three dust 
condensations labelled as Sgr B2(N), (M) and (S).
They contain all the tracers of on--going star
formation: ultracompact HII regions driven by the UV field of newly born
$OB$ stars, hot cores embedding proto--stars,
molecular masers and high far--IR continuum intensity.
These \textit{core} regions are surrounded by a 
low density (n$_{H_2}$$\leq$10$^4$ cm$^{-3}$) extended envelope 
($\sim$15$'$), hereafter Sgr B2 envelope, of warm gas (T$_k$$\geq$200 K) 
and cool dust (T$_d$=20-30 K; H\"{u}ttemeister et al. 1995). 
A summary of the different components  present in Sgr B2
and their main characteristics is shown in  figure \ref{fig:1} [{\textit{left}}]. 

The origin of the observed rich chemistry in the Sgr B2
envelope and its possible heating mechanisms are far
from settled and several scenarios have been proposed.
Low--velocity shocks have been traditionally invoked to explain the enhanced
abundances of SiO or NH$_3$ and the differences between
gas and dust temperatures (cf. Flower et al. 1995).
The origin of  shocks in the Sgr B2 envelope have
been associated either with large scale cloud--cloud collisions
or with small scale wind--blown bubbles produced by evolve massive stars
(Mart\'{\i}n--Pintado et al. 1999).

The effect of the UV radiation in the Sgr B2 envelope has been traditionally
ruled out because of the gas and dust temperature differences, the
unusual chemistry and the abscense of thermal radio-continuum and ionized gas
outside the HII regions and hot cores within the central condensations.
Our observations reveal the presence  of an extended component of
ionized gas detected by fine structure emission.

All this new data suggest that UV radiative--type processes are
also important in the heating of the GC gas in addition
to mechanical mechanisms, as can be in other GC clouds
(see Rodr\'{\i}guez-Fern\'andez et al. 2003). 
In this contribution we present a
brief summary of the results obtained by ISO\footnote{Based on observations 
with ISO, an ESA project with instruments funded by ESA Member States 
(especially the PI countries: France, Germany, the Netherlands 
and the United Kingdom) and with participation of ISAS and NASA.} in 
the Sgr B2 envelope (Fig. \ref{fig:1} [\textit{right}]) 
concerning the ionized gas and the effects of the UV radiation.

\section{Extinction corrections}

The large H$_2$ column density (up to 10$^{25}$ cm$^{-2}$) found in 
Sgr B2 suggests that even in the far--IR, fine structure lines can suffer
appreciable reddening. 
We have estimated the prevailing extinction
in each position by converting the continuum opacity into visual
extinction. The spectra can not be fitted with a single gray body.
Thus, we have modeled the observed continuum
spectrum as a sum of two gray bodies. 
The total continuum flux in the model is:
\begin{equation}
  S_\lambda \; = \;
(1-e^{-\tau_\lambda^{warm}})\;B_\lambda(T_{warm})\;\Omega_{warm}
+ (1-e^{-\tau_\lambda^{cold}})\;B_\lambda(T_{cold})\Omega_{cold}    
\label{eq-flujoBB}
\end{equation}  
where B$_\lambda$(T$_i$) is the Planck function, $\tau_\lambda^i$ is the continuum
opacity and $\Omega_i$ is the solid angle subtended by the $i$ dust component.
The continuum opacity is given by 
$\tau_\lambda \; = \; 0.014\;A_V\;\left(30/\lambda\right)^\beta$
where $\beta$ is the grain emissivity exponent of each dust component.
The observed spectral energy distributions are
best fitted  with a dust component with a temperature of 13--22 K and
a warmer component with a temperature of 24--39 K.
The warmer component contribute in less than 10 \%  to the total
optical depth. The higher dust temperatures are those measured in the
southern part of Sgr B2. Depending of the position, the derived
extinction varies from $\sim$20 to $\sim$1000 magnitudes 
(see table \ref{ratios} for  lower and upper limits to the
visual extinction).

\section{The ionized gas}

The far--IR spectra exhibit several fine structure lines from the
ionized material. 
We have clearly detected  the [OI]63 and 145 $\mu$m, and [CII]158 $\mu$m
lines in all observed positions. In addition, lines coming from 
the [NII]122 $\mu$m, [NIII]57 $\mu$m [OIII]52 and 88 $\mu$m lines are 
also detected in most positions, revealing  a prominent
component of ionized gas in the southern and eastern 
regions of Sgr B2.

Table 1 lists the extinction corrected OIII 52/88 line
intensity ratios for the two derived limits to the extinction across the region. 
For extended emission sources
and lines excited by collisions with electrons  (see Rubin et al. 1994)
we derive electron densities (n$_e$) in the range $\simeq$10$^{2-3}$ cm$^{-3}$
for the extended envelope. 
At the limited spectral resolution of the
LWS/grating mode, OIII lines are hardly detected in the central positions.   
Nevertheless, Fig. 2b shows their  Fabry--Perot detection
in Sgr B2(M). 
Both OIII lines appear centered at V$_{LSR}$$\simeq$50$\pm$15 km s$^{-1}$ and
do not show emission/absorption at more negative velocities (foreground
gas). Considering A$_V$$>$1000 magnitudes, 
we found n$_e$$\geq$10$^{5.5}$ cm$^{-3}$ in Sgr B2(M).

Table 1 also lists the extinction corrected NIII57/NII122 line intensity 
ratios. The minimum averaged ratio is $\sim0.77$ while the upper limits are 
dependent to the maximum extinction affecting the  lines.
For those ratios, we derive minimum effective temperatures (T$_{eff}$) 
for the ionizing radiation of $32000-36000\;K$. 
Those T$_{eff}$ should be considered as a lower limit to
the actual T$_{eff}$ of the ionizing source if this is
located far from the nebular gas.
We have carried out  CLOUDY (Ferland 1996) simulations showing that 
the observed line ratios are consistent with an scenario where almost
all ionizing photons arise from the HII regions  within  Sgr B2(M)
and (N).
The total flux of Lyman continuum photons  is 
10$^{50.3}$ s$^{-1}$ and T$_{eff}=35000\;K$. The differences in the observed  NIII/NII ratios
are due to the dilution of the incident radiation (lower ionization
parameter). Hence, the size of the ionized region can only be explained if
the medium is highly inhomogeneous.
This suggests that the clumpy nature of the cloud allows the radiation to 
reach large
distances through the envelope. In this scenario, several PDRs can be
expected in the interface between the ionized and the neutral gas.

\begin{table}[hb]
\caption{Selected line ratios after correcting for the estimated minimum
and maximum extiction. The different beam sizes of each LWS
detector are taken into account and extended emission is considered.
Offsets are in arsec.} 
\label{ratios}\renewcommand{\arraystretch}{0.8} 
\begin{tabular}{cccccccc}\hline

\small map      & \small warm dust   & \small cold
dust    & \small OIII     & \small n$_e$(OIII)    &
\small NIII/NII & \small T$^{min}_{eff}$ & \\
\small position & \small A$_V$ (mag) & \small A$_V$ (mag)  & \small
R(52/88) & \small log(cm$^{-3}$) & \small R(57/122)& \small ($\div10^3\;K$) & \\
\hline
\small (0,810) & \small 1.2-1.5  & \small 15-55   & \small  $<$2.40   & \small $<$3.18   & \small $<$0.72   & \small $<$33.2 &\\
\small (0,630) & \small 1.6-2.0  & \small 23-84   & \small  1.79-2.23 & \small 2.83-3.10 & \small $<$0.72   & \small $<$33.2 &\\
\small (0,450) & \small 1.1-1.8  & \small 25-92   & \small  1.36-1.73 & \small 2.49-2.79 & \small $<$0.26   & \small $<$31.8 &\\
\small (0,270) & \small 5.2-8.0  & \small 41-112  & \small  $<$1.61   & \small $<$2.70   & \small $<$1.77   & \small $<$35.6 &\\
\small (0,180) & \small 0.7-1.7  & \small 131-294 & \small  $<$3.55   & \small $<$3.66   & \small $<$2.73   & \small $<$35.2 &\\
\small (0,-90) & \small 16-18    & \small 367-877 & \small  1.27-8.11 & \small 2.41-4.67 & \small 1.60-11.3 & \small (35.0-37.2)&\\
\small (0,-180)& \small3.7-5.2  & \small 148-493 & \small  0.90-3.18 & \small 1.99-3.53 & \small 0.78-2.95 & \small (34.9-35.3) &\\
\small (0,-270)& \small2.8-3.8  & \small 59-205  & \small  0.41-0.69 & \small 1.03-1.67 & \small 0.79-1.39 & \small (35.0-35.8) &\\
\small (0,-450)& \small2.7-3.2  & \small 28-102  & \small  1.14-1.50 & \small 2.28-2.61 & \small 0.18-0.24 & \small (31.6-32.2) &\\
\small (0,-630)& \small3.5-3.9  & \small 23-85   & \small  1.83-2.30 & \small 2.85-3.13 & \small 0.55-0.71 & \small (32.9-33.2) &\\
\small (0,-810)& \small1.1-1.3  & \small 16-59   & \small  0.55-0.64 & \small 1.39-1.58 & \small 0.18-0.21 & \small (32.8-33.1)  &\\ 
\small (270,0) & \small 27-28    & \small 156-536 & \small  $<$1.19   & \small $<$2.24   & \small $<$3.13   & \small $<$36.6 &\\ 
\small (180,0) & \small 4.9-6.6  & \small 78-276  & \small  1.14-2.38 & \small 2.28-3.18 & \small $<$0.95   & \small $<$33.6 &\\ 
\small (90,0)  & \small 7.4-9.0  & \small 168-565 & \small  0.62-2.64 & \small 1.53-3.30 & \small 1.31-6.07 & \small (35.7-36.3) &\\
\small (-90,0) & \small 28-34    & \small 228-579 & \small  0.85-2.97 & \small 1.92-3.44 & \small           & \small &\\
\small (-180,0)& \small21-26    & \small 62-168  & \small  0.43-0.63 & \small 1.10-1.55 & \small           & \small &\\
\small (-270,0) & \small 21-25   & \small 45-124
& \small  0.38-0.50 & \small 0.94-1.27 & \small           & \small &\\
\hline
\end{tabular}
\end{table}
	
\clearpage

\begin{figure}[ht]
\begin{center}
\includegraphics[height=17cm]{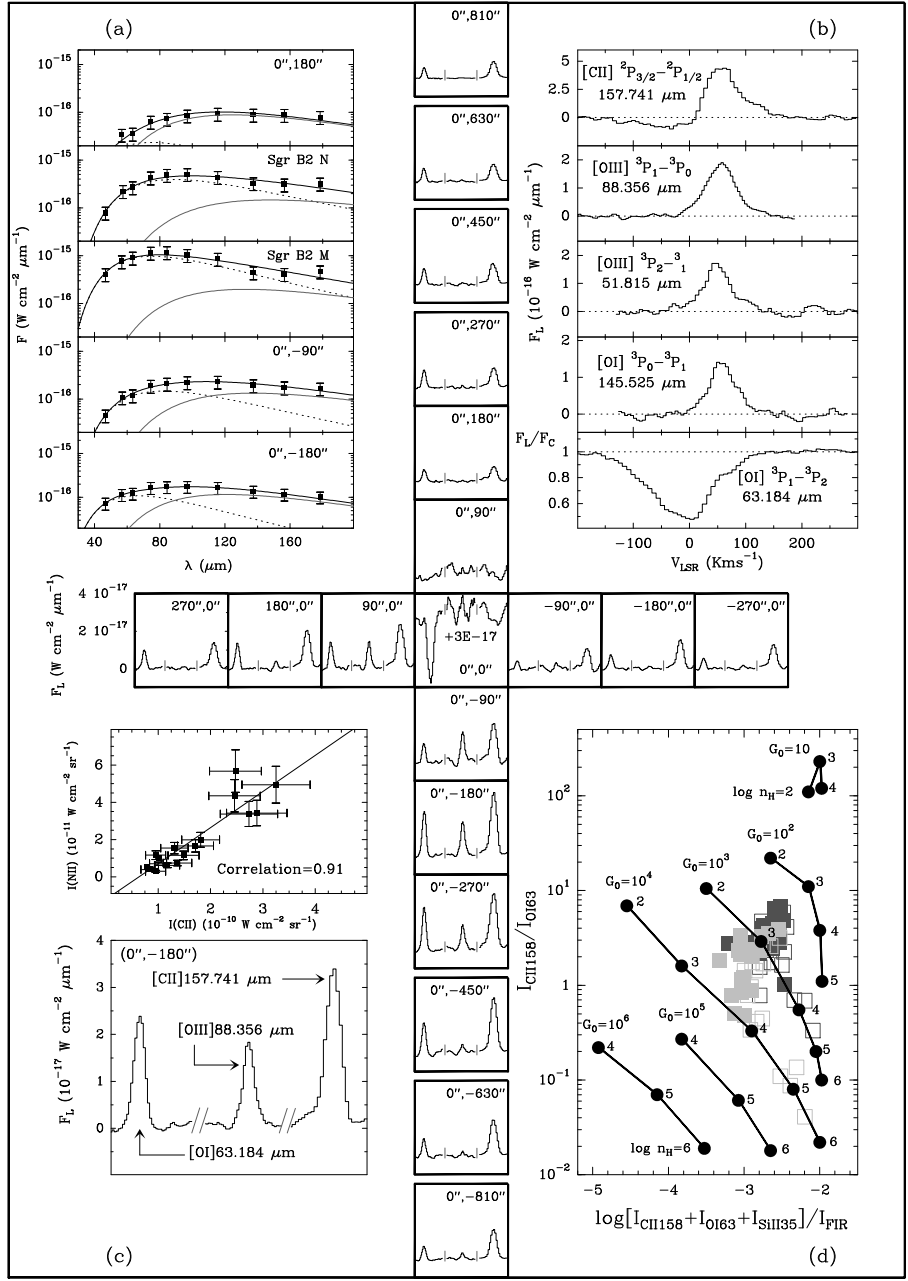}
\end{center}
\caption{ISO/LWS--grating mode raster map  of the
[OI]63, [OIII]88 and [CII]158 $\mu$m lines. Offset positions
are given in arcsec respect to Sgr B2(M) $(0'',0'')$.
(a) Averaged continuum flux of each ISO/LWS detector and gray--body
best fits for some selected positions. The error bars correspond to a
30 \% of flux uncertainity. 
(b) Fine structure lines detected with ISO/LWS--FP mode in Sgr B2(M).
(c) [\textit {Top}] Correlation between the [NII]122 $\mu$m and 
[CII]158 $\mu$m lines. SgrB2(M) and (N)  are not included . 
[\textit {Bottom}] Main features of the grating raster map labelled. 
(d) CII/OI intensity ratio vs. (CII+OI+SiII)/FIR for several
PDR models of varying FUV fields and hydrogen densities (from 
Wolfire et al. 1990). The space parameter ocuppied by 
Sgr B2 envelope is represented by gray squares (see text).}
\label{fig:2}
\end{figure}

\clearpage

\section{The Photodissociation regions}

The large OH column density of warm (T$_k$$\simeq$300 K) and low density gas
(n$_{H_2}$$\leq$10$^4$ cm$^{-3}$) recently found in the envelope of Sgr B2 
suggested that its external shells are illuminated by a
strong far--UV field that photodissociate the large amount of water vapor
found in the region (Goicoechea \& Cernicharo 2002). However, the
main properties of the associated warm PDRs could not be derived.
For that porpose, we have compared the far--IR continuum emission, and 
the [CII] 158, [OI]63 and 145 $\mu$m lines at each position 
with PDR theoretical models. 
Figure 2d shows our results in comparison with predictions from
PDR models of  Wolfire, Tielens and Hollenbach (1990).
The gray squares show the parameter space occupied
by the Sgr B2 positions for intensity ratios corrected for the minimum 
(filled) and maximum (not filled) visual extinction estimated from
the continuum analysis.
The observational points scatter around a far--UV flux, G$_0$, of
$\simeq$10$^{3-4}$ and
n$_H$$\simeq$10$^{3-4}$ cm$^{-3}$ depending on the use of all observed 
CII flux (dark gray) or the remaining CII flux  after subtracting the CII 
emission arising in (non--PDR) low--density ionized gas and
correlated with the NII emission.
From the observed  correlation (see Fig. 2c) we estimate that
20 to 70 \% of the CII emission arises in the PDRs of Sgr B2.    
In addition, both the observed [CII]158/[OI]63 and ([CII]158+[OI]63)/FIR
intensity ratios shown in Fig. 2d are not predicted  by shocked gas models
(Hollenbach \& McKee 1989) while they are commonly reproduced in 
PDR models.

\section{Summary and perspectives}

We have presented new far--IR observations of the Sgr B2 region
that reveal a new perspective of the less known extended envelope 
of the complex. The ISO data show the presence of a widespread component
of ionized gas reaching very large distances from the HII regions of
known massive star formation. Molecular tracers and atomic fine structure
tracers do not show evidences
of high--velocity shocks. Hence, the ionized gas can not be explained
in terms of high--velocity dissociative shocks. 
It seems that the well established
widespread low--velocity shocks are not the only mechanism heating the gas to
temperatures larger than those of the dust.
We now have proofs showing that low-velocity shocks and large scale 
radiative processes, such as the photoelectric effect, coexist in the whole 
Sgr B2 envelope, difficulting the
interpretation of the astronomical data but providing one of the richest 
and peculiar clouds in the galaxy.
Coexistence of mechanical and radiative--type  heating mechanisms based on the
effects of a far--UV field that permeates an inhomogeneous and clumpy medium
seems to be the rule in the envelope of Sgr B2 and in the bulk of GC 
clouds observed by ISO (Rodr\'{\i}guez-Fern\'andez et al. 2003).

\begin{acknowledgement}
We thank J. Mart\'{\i}n--Pintado for stimulating discussions about 
Sgr B2 and the Galactic Center, and M.A. Gordon for providing us the IRAS 
maps of Sgr B2.
NJR-F has has been supported by a Marie Curie Fellowship of the European
Community program ``Improving Human Research Potential and the Socio-economic
Knowledge base'' under contract number HPMF-CT-2002-01677.
\end{acknowledgement}


\end{document}